\documentclass[noshowpacs,amsmath,twocolumn,
aps,prb]
{revtex4-1}

\usepackage[left]{lineno} 
\usepackage{siunitx}  
\usepackage{setspace}
\usepackage{amsmath}
\usepackage{graphicx}
\usepackage[nearskip,margin = 0pt]{subfig}
\usepackage{subfig}

\usepackage{verbatim}
\usepackage{amsfonts}
\usepackage{amssymb}
\usepackage{epstopdf} 
\usepackage{xcolor}
\usepackage{color}
\DeclareGraphicsExtensions{.pdf,.eps,.png,.jpg,.mps} 
\usepackage{ragged2e}
\usepackage{hyperref}
\hypersetup{
    colorlinks=true,
    linkcolor=blue,
    citecolor=blue,
    filecolor=blue, 
    urlcolor=blue,
}
\usepackage{mathrsfs}
\usepackage{float} 

\begin{document}
\title{High-brightness multimode fiber laser amplifier}

\author{ Zhen Huang$^{1,2,\dag}$, Binyu Rao$^{1,2,\dag}$,  Zefeng Wang$^{1,2,\dag,*}$, Chenxin Gao$^{1,2,\dag,*}$, Hu Xiao$^{1,2}$, Bokai Yi$^{1,2}$, Zilun Chen$^{1,2}$, Pengfei Ma$^{1,2}$, Jiajia Zeng$^{1,2}$, Dongran Shi$^{1,2}$, Baolai Yang$^{1,2}$, Xiaofei Ma$^{1,2}$, and Xiangfei Zhu$^{1,2}$, \\
$^1$ College of Advanced Interdisciplinary Studies, National University of Defense Technology, Changsha, China.\\
$^2$ Nanhu Laser Laboratory, National University of Defense Technology, Changsha, 410073, China.\\ 
${\dag}$ These authors contributed equally to this work.\\ 
${*}$ Corresponding authors: zefengwang$\_$nudt@163.com. chenxingao24@nudt.edu.cn.}

\maketitle
\newcommand{\ts}{\textsuperscript}
\newcommand{\tsb}{\textsubscript}

\vspace{300 mm}

\noindent{\bf Abstract} 

Fiber lasers are widely used in various fields owing to their high efficiency, flexible transmission and excellent beam quality. In applications such as industrial manufacturing and defense systems, a higher output power is always desired. Nevertheless, the power scaling in fiber lasers is limited by nonlinear effects and transverse mode instability in conventional high-power fiber laser systems, where the laser is amplified within the fundamental fiber mode. A promising strategy to overcome these limitations is to utilize multimode fibers, which exhibit higher thresholds for both nonlinear effects and transverse mode instability, combined with wavefront shaping techniques to convert the output speckle pattern into a single concentrated spot.
In this study, a high-power multimode fiber laser amplifier based on wavefront shaping is constructed and investigated, achieving a focused beam profile with a 168 W output power. The effects of objective function and the linewidth of seed laser on the system performance are also studied. Additionally, an all-fiber version of high-brightness multimode fiber laser amplifier is proposed. This work opens up new avenues for leveraging multimode fibers to achieve higher brightness in fiber lasers and may inspire other research based on wavefront shaping.

\vspace{3 mm}
\noindent{\bf Introduction} 

Fiber lasers have extensive applications in communication, sensing, advanced manufacturing, and scientific research. In some applications such as industry and defense, high-power and especially high-brightness fiber lasers are demanded\cite{12high_power_fiber_lasser_review_Science_2011}. Since the first demonstration of double-clad fiber in 1988\cite{first_double_cladding_fiber_1988}, the power of fiber lasers has scaled at an exceptional pace\cite{11high_power_fiber_lasser_review_NP_2013}. However, contrast to the rapid progress observed before 2012, the scaling of output power for high-beam-quality fiber lasers has plateaued over the past decade\cite{rby8,rby9}, which is primarily attributed to the limitations imposed by nonlinear effects and transverse mode instability (TMI)\cite{TMI_review_2020,rby10}. Nonlinear effects lead to wavelength conversion or backward signal propagation during laser amplification\cite{rby11}, while TMI induces dynamic coupling between the fundamental and higher-order fiber modes, leading to the beam quality deteriorating\cite{TMI_light}. 

In high-power fiber laser systems, the fundamental fiber mode has traditionally been prioritized to preserve high beam quality, with single-mode or few-mode fibers commonly employed. Moreover, most of the studies addressing the aforementioned limitations have been restricted to fundamental mode amplification. Although this ensures that the brightness will increasing almost linearly with the power, existing nonlinear and TMI suppression methods based on fundamental mode output have relatively limited effectiveness. Recent studies have shown that when multiple modes are excited in multimode fibers (MMFs), the stimulated Brillouin scattering (SBS) and TMI thresholds are increased by roughly an order of magnitude, compared to conventional fundamental mode excitation \cite{Further-Theoretical-Study-on-High-TMI-Threshold-in-Multimode-Systems_APL_2024,High-TMI-Threshold-in-Multimode-Systems_PNAS_2023,High-SBS-Threshold-in-Multimode-Systems_NC_2023,Optimal_Excitation_for_Improving_TMI_and_SBS_Thresholds_in_MMF_2024}. In TMI, the diffusive nature of heat propagation causes the transverse homogenization speed of the temperature field (or the decay rate of temperature eigenmodes) increasing with higher transverse spatial frequencies. This leads to weak thermal-optical coupling between non-adjacent modes, forming a sparse modes coupling matrix. Due to the sparsity of the matrix, the effective coupling power obtained by each mode is relatively small, thereby increasing the overall TMI threshold\cite{Further-Theoretical-Study-on-High-TMI-Threshold-in-Multimode-Systems_APL_2024,High-TMI-Threshold-in-Multimode-Systems_PNAS_2023}. In SBS, multimode excitation can effectively broaden the Brillouin gain spectrum. This broadening arises from frequency shifts discrepancies among different forward and backward modes pairs in Brillouin scattering, which reduces the peak gain, ultimately leading to an increased SBS threshold\cite{High-SBS-Threshold-in-Multimode-Systems_NC_2023}. These findings provide a new direction to break the current limitations and further enhance the power of fiber lasers. 

\begin{figure*}[th!]  
\includegraphics[width=0.95\linewidth]{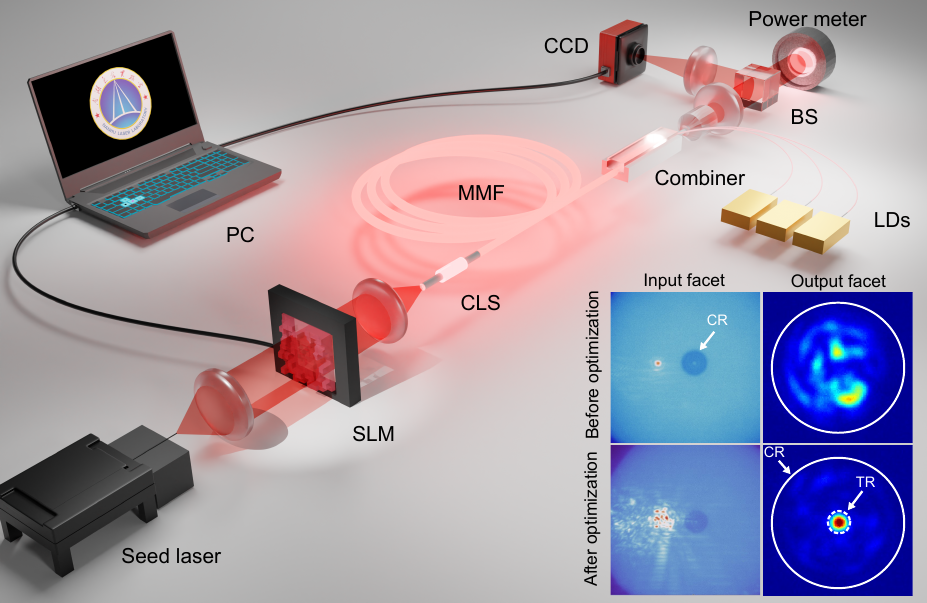}  
\captionsetup{singlelinecheck=off, justification=centerlast}  
\caption{{\bf Schematic diagram of the high-brightness multimode fiber laser amplifier.} The collimated seed laser is modulated by a spatial light modulator (SLM) before being focused into the MMF. The backward-propagating pump laser is introduced via a pump-signal combiner. A cladding light stripper (CLS) is employed to eliminate the residual pump laser as well as forward-propagating unwanted cladding laser from the seed. The beam profile at the output facet of MMF is imaged through a charge-coupled device (CCD), providing feedback to the SLM. The insets depict representative experimental beam profiles at the input and output facets of the MMF before and after optimization. The MMF utilized in these insets is characterized by a core diameter of 50 $\si{\micro\meter}$ and a NA of 0.12. When capturing the beam profile at the input facet, the beam spot was deliberately displaced out of the core to illustrate its dimensions relative to the fiber core.}  
\label{fig1}  
\end{figure*}

Although MMFs provide enhanced nonlinear thresholds, their multimodal characteristics introduce new challenges. The coherent superposition of multiple modes generally manifests as a speckle, which is undesirable for applications that require high-brightness lasers. The brightness of a laser beam is proportional to its power $P$, while inversely proportional to the square of the product of spot radius $r$ and angle of divergence $\theta$ : 
\begin{equation}
\begin{aligned}
B\propto \frac{P}{\left( r\times \theta \right) ^2}
\label{eqn1}
\end{aligned}
\end{equation}
The output power is ensured by the high nonlinear threshold of MMFs and the divergence angle of the output beam is limited by the numerical aperture (NA) of the fiber core. Consequently, achieving high-brightness output in fiber lasers requires minimizing the spot size at the output facet of the MMF. Fortunately, provided that modal coherence is preserved, a high-brightness, tightly focused laser spot at the fiber output facet can be achieved through wavefront shaping (WFS) technique\cite{Compressed_Sampling_Method_for_WFS_Transmission_Matrix_Measurement_2021,Feedback_Based_WFS_Overview_2015,First_Proposal_of_WFS_2007,Internal_Focusing_in_Scattering_Media_NP_2022,Liu_Qiang_paper,NP_Overview_of_Light_Field_Modulation_in_Complex_Media_2022,Proposal_of_GA_in_WFS_2012,Proposal_of_WFS_Transmission_Matrix_Method_2010,WFS_Multimode_Fiber_Far_Field_Optimization_2024,Spatiotemporal_Shaping_in_Complex_Media_NP_2024}. This method effectively controls the modes excited at the input facet of the MMF, enabling researchers to manipulate the beam profile at the output facet. For an output beam at the fiber facet satisfying the condition \(\text{NA} \times \omega_0 \approx \frac{\lambda}{\pi}\), where \(\omega_0\) is the radius of the spot, the brightness approximates that of a Gaussian beam with the same power\cite{Siegman_Lasers_1986}. Actually, WFS has already been demonstrated to effectively manipulate the output beam of MMF even in the gain-saturated\cite{Verification-of-WFS-Effects-in-Gain-Multimode-Fibers_Light_2016,WFS_in_gain_multimode_based_on_transfer_matrix_2019,phase_conjugate_WFS_in_multicore_fiber_2010,Theoretical_Study_of_Multimode_Amplifiers_Based_on_WFS_2024} or nonlinear\cite{WFS-Control-of-Nonlinearity-in-Multimode-Fibers_2018,Multidimensional-Lasers-in-Spatiotemporal-Mode-Locked-Cavities-with-WFS_2020} regime. Integrating the beam control ability of WFS with the high threshold of nonlinear and TMI effects in MMFs provides a new strategy for enhancing the brightness of fiber lasers. Although the promising prospects of MMFs in high-power fiber lasers has been predicted in previous studies \cite{Verification-of-WFS-Effects-in-Gain-Multimode-Fibers_Light_2016,High-SBS-Threshold-in-Multimode-Systems_NC_2023, Verification-of-WFS-Effects-in-Gain-Multimode-Fibers_Light_2016,WFS_in_gain_multimode_based_on_transfer_matrix_2019}, there have been no detailed research in WFS-based high-power multimode fiber laser amplifiers and its performance has not been systematically investigated.

In this work, we constructed a high-power multimode fiber laser amplifier based on WFS. The schematic diagram of our system is presented in Fig.~\ref{fig1} along with the typical beam profiles on the input and output facets of the MMF before and after optimization. A hundred-watt-level focused beam profile is achieved. The influences of the objective function in WFS and the linewidth of the seed laser on the performance of the system were investigated systematically through simulations and experiments. Furthermore, our results indicate that an all-fiber version of the high-brightness multimode fiber laser amplifier can be realized by intermodal phase control. These findings demonstrate that MMFs, employed as gain media in fiber laser amplifiers, are capable of being controlled in high-power amplification regime, which opens up the avenue for exploiting MMFs to break the current power limitations in fiber lasers.

\vspace{3 mm}
\noindent{\bf Result}

\noindent{\bf 1.WFS in multimode fiber laser amplifier}

\begin{figure*}[th!]
\includegraphics[width=0.95\linewidth]{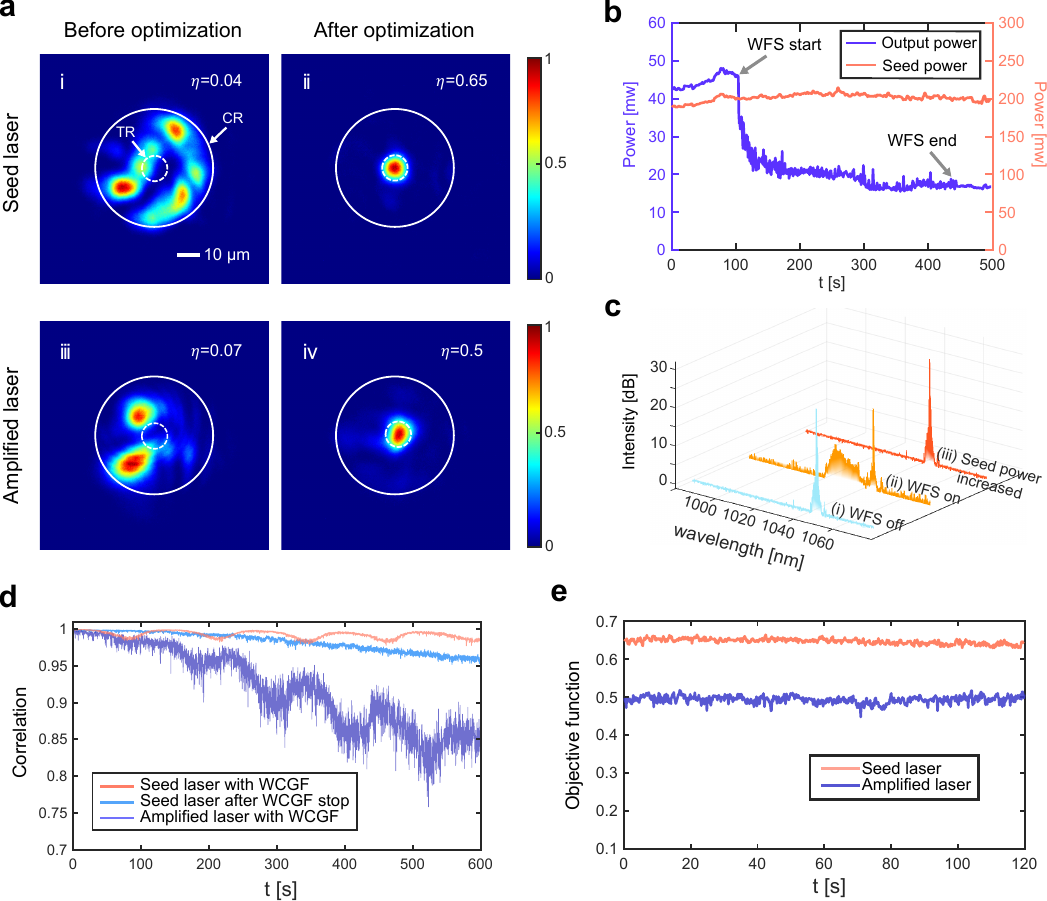}
\centering
\captionsetup{singlelinecheck=off, justification =centerlast}
\caption{{\bf High-power experimental results of the multimode fiber laser amplifier base on WFS.} {\bf a} Beam profiles before and after optimization. In the beam profiles, TR and CR are circled with dashed and solid lines respectively. The diameter of the TR was set to 12 $\si{\micro\meter}$, with the final objective functions reaching 0.65 and 0.5 for seed laser and amplified laser, respectively. {\bf b} Evolution of seed power and output power during the optimization process of WFS with seed laser injection only. Before the optimization, the output power was approximately 20\% of the seed power, which is caused by the absorption of gain fiber. {\bf c} Output optical spectra of the multimode amplifier. (i) illustrates the output spectrum when the seed laser was amplified from 10 W to 170 W without WFS. When WFS was applied to seed laser, ASE appeared and the output spectrum corresponds to (ii). Subsequently, the seed power was increased to 18 W and the ASE disappeared, as shown in (iii). {\bf d} The stability of the output beam profile under different conditions. The output beam profiles were recorded during a period of time and their correlations were calculated under three conditions respectively: (1) seed laser injection only with the water cooling system of the gain fiber (WCGF) turned on (orange), (2) seed laser injection only, after the moment of WCGF turned off (blue), and (3) seed laser amplification to 170 W with the WCGF turned on (purple). {\bf e} Stability of the objective function during a period of time after optimization with (purple) and without (orange) pump.
}
\label{fig2}
\end{figure*}
As illustrated in Fig.~\ref{fig1}, a linearly polarized single-frequency seed laser operating at 1045 nm is phase-modulated by a phase-only spatial light modulator (SLM) and then coupled into a MMF for amplification. The beam profile, optical power and spectrum of the amplified laser are detected by the charge-coupled device (CCD), power meter and optical spectrum analyzer (OSA), respectively. The fiber laser amplifier employs a MMF with a core diameter of 48 $\si{\micro\meter}$ and a NA of 0.06, supporting 21 modes. The SLM plays a pivotal role in WFS, adjusting the mode composition and intermodal phase of the seed laser at the input facet of the MMF. Genetic algorithm (GA) is employed as a WFS method, which has been proven robust against noise and perturbations \cite{Proposal_of_GA_in_WFS_2012}. The SLM is divided into 16 $\times$ 16 macro pixels to reduce the degree of freedom of optimization. The objective of optimizing the phase map on the SLM is to confine the output beam within a small circular region, thereby achieving high brightness in the output laser. The optimization objective function, $\eta$, is defined as:
\begin{equation}
\begin{aligned}
\eta =\frac{\iint_{\mathrm{TR}}{I(x,y)\mathrm{d}x\mathrm{d}y}}{\iint_{\mathrm{CR}}{I(x,y)\mathrm{d}x\mathrm{d}y}}
\label{eqn2}
\end{aligned}
\end{equation}

where CR is the core region on the output facet of the MMF and the TR (target region) is a circular region in the fiber core. The diameter of TR is set to 12 $\si{\micro\meter}$, corresponding approximately to the diffraction-limited spot size for a NA of 0.06 (Detailed experimental configurations are described in the Method Sec.~1-2 and Supplementary Sec.~1).

The coupling efficiency of seed at the input facet of MMF plays an important role in fiber laser amplifier systems, which has been overlooked in prior studies of WFS in MMF. Consequently, we firstly conducted WFS with the seed laser injection alone before proceeding to the high-power amplification experiments. The coupling efficiency of the modulated seed laser is evaluated to ensure adequate seed power in high-power amplification regime. Additionally, the optimization performance without gain is investigated as a comparison to the high-power amplification result. The beam profiles before and after optimization are shown in Fig.~\ref{fig2}a(i) and (ii) respectively. After optimization, the objective function is 0.65. Variation of seed power and output power of the MMF during the optimization process are presented in Fig.~\ref{fig2}b. At the initiation of WFS, the output power of seed laser decreased dramatically by approximately half compared to the initial state. In high power regime, the decrease in seed power does not considerably reduce output power but may lead to significant amplified spontaneous emission (ASE), which may pose a risk to the system. Adopting larger macro-pixels in the gene encoding scheme of GA can mitigate the degradation of coupling efficiency induced by WFS; however, this inherently reduces the degrees of freedom available for optimization, which compromises the control ability of output beam, manifesting as a rapid convergence of objective function to a suboptimal value. In practice, there is a trade-off between the degrees of freedom for optimization and the coupling efficiency of seed laser.

Subsequently, we investigated the performance of WFS in the high-power amplification regime. The seed power was increased to 10 W and the pump power was raised to approximately 212 W. The seed laser was amplified to 170 W, achieving an amplification efficiency of 80\%. The output spectrum exhibited a clean signal peak, as shown in the curve of Fig.~\ref{fig2}c(i). The SLM was then activated to optimize the wavefront of the seed laser. In the high-power amplification regime, the output power is nearly impervious to reduction in coupling efficiency of seed laser. Instead, the reduction in coupling efficiency manifests as ASE in the output optical spectrum. When WFS was activated, the output power decreased to 160 W, and a pronounced ASE peak appeared in the spectrum, as shown in  Fig.~\ref{fig2}c(ii).  To address this, the seed power was increased from 10 W to 18 W, which restored the spectral shape (Fig.~\ref{fig2}c(iii))  and maintained stability throughout the subsequent optimization process. 

In the high-power amplification regime, the beam profiles before and after optimization are presented in Fig.~\ref{fig2}a(iii) and (iv). After optimization, the objective function is 0.5, with a 168 W output power. This indicates that the brightness after WFS is approximately 8 times greater than the uncontrolled state. The objective function is lower than WFS without pump, which could be attributed to the intensified beam profile fluctuations in the high-power amplification regime. Because our goal is to achieve high brightness rather than specific spot positioning, and during the final stages of optimization, the focused spots initially deviated from TR but gradually migrated toward it through algorithm iterations. Consequently, to accelerate the convergence of the objective function, the position of TR was allowed to shift adaptively in final stage of optimization. The modified approach focused on enhancing the power ratio within the current focused spot, rather than strictly repositioning the spot to the center of fiber core.

\begin{figure*}[th!]
\includegraphics[width=0.95\linewidth]{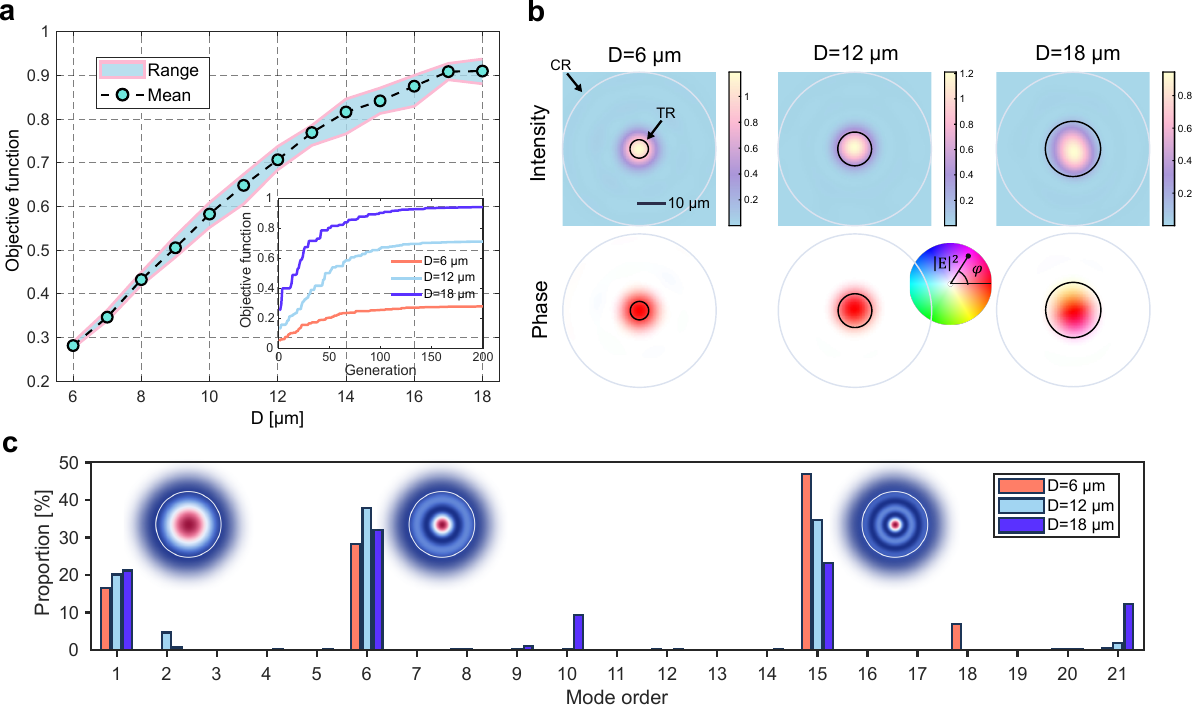}
\centering
\captionsetup{singlelinecheck=off, justification = centerlast}
\caption{{\bf Simulation results of the superposition of multiple fiber modes to achieve a diffraction-limited spot.} {\bf a} The optimized Objective functions for different diameters of TR. The TR is selected at the center of the fiber core. For each diameter D, the algorithm was executed 10 times, and ranges and average values of the Objective functions are presented. The inset shows the changes of objective functions for D=6 $\si{\micro\meter}$, D=12 $\si{\micro\meter}$, D=18 $\si{\micro\meter}$. {\bf b} The Intensity and phase distributions after optimization for D=6 $\si{\micro\meter}$, D=12 $\si{\micro\meter}$, D=18 $\si{\micro\meter}$ respectively.  {\bf c} Mode proportion after optimization for D=6 $\si{\micro\meter}$, D=12 $\si{\micro\meter}$, D=18 $\si{\micro\meter}$. The simulations were conducted utilizing a MMF with core diameter of 48 $\si{\micro\meter}$ and a NA of 0.06, which supports 21 modes(3 radial modes and 18 angular modes). The diameter of the diffraction-limited spot corresponding to the fiber is approximately 11 $\si{\micro\meter}$.
}
\label{fig3}
\end{figure*}

The stability of the system was also analyzed. Achieving a concentrated beam profile through mode combining requires a stable mode proportion and phase. The mode proportion is considered to be stable before mode instability occurs. Furthermore, the random mode coupling is negligible considering that the length of the MMF used in this work is just few meters. Therefore, intermodal phase perturbation is much more critical to achieve stable output and control the output beam profile. We characterized the stability of the output beam profile by calculating the correlation between the first frame and subsequent frames before optimization (see Method Sec.~4 for definition of the correlation). The correlation of the beam profile remains close to unity under ambient conditions when the pump was off. And the activation of the water cooling system of the gain fiber (WCGF) induces periodic fluctuations in beam profile (Fig.~\ref{fig2}d). This behavior is mainly caused by the periodic temperature oscillation of the WCGF around its setting temperature. Moreover, when the WCGF is deactivated, allowing the gain fiber to gradually equilibrate to room temperature, the output beam profile evolves progressively. The noise observed in the curve are attributed to tiny perturbations in the system. In the high-power amplification regime, both the periodic variations and noise in the curve escalate markedly. The stronger periodic variations reflects a intensified temperature oscillation within the WCGF due to the increased thermal load. And the higher level of noise suggest that they are not solely caused by external mechanical disturbances. Instead, thermal-optical effect within the fiber could be a significant contributing factor and amplify the disturbances, which is explained in the previous simulation results\cite{Theoretical_Study_of_Multimode_Amplifiers_Based_on_WFS_2024}. The objective functions after optimization were recorded over 2 minutes, as presented in Fig.~\ref{fig2}e. It is proven that the objective function remains stable after WFS under both seed laser transmission condition and high-power amplification regime.

\vspace{3 mm}
\noindent{\bf 2.Effect of TR on the optimization performance}

The electric fields of the fiber modes can be regarded as an incomplete orthogonal function basis due to the limited number of supported modes. This means that not all electric field distributions can be synthesized by mode superposition. To determine theoretically the maximum power ratio within TR for a given type of fiber, the focusing capability of the output beam profile for a given fiber is investigated by simulation, which is analogous to evaluating the temporal compression of a pulse within a finite frequency bandwidth.

The simulation is conducted by searching various kinds of modes superpositions to optimize the beam profile utilizing GA with different TR sizes (see the Method Sec.~3 for details). For each TR size, the optimization was executed 10 times to mitigate the stochasticity of GA. Single optimization converges well (inset of Fig.~\ref{fig3}a) and there is little variation among multiple optimizations, indicating that GA can reliably identify the global optimum of the defined objective function. As illustrated in Fig.~\ref{fig3}a, the ratio of power within the circle obtained by mode superposition increases as the TR enlarges. However, when the diameter of TR reaches 18 \textmu m, the intensity and phase distribution are distorted (Fig.~\ref{fig3}b), which indicates that the beam profile deviates from the diffraction-limited spot. When the TR is positioned at the center of the fiber, the optimized output is dominated by radial modes, with proportion varying depending on TR size (Fig.~\ref{fig3}c). By altering the size or position of the TR and performing multiple GA, it is observed that employing this intensity-dependent objective function uniquely determines the optimized mode proportion and phase. However, if the TR of the objective function is excessively large comparing to the corresponding diffraction-limited spot, it will lead to the distortion of the focused spot and increase the uncertainty of the output mode proportion and phase. In contrast, too small TR will spread power into the background across the entire CR. Therefore, it is necessary to make the size of TR close to that of the diffraction-limited spot corresponding to the fiber. Essentially, defining the objective function as the power ratio within a small circle is not simply to concentrate the intensity distribution, the more fundamental purpose is to achieve a smooth phase distribution. Actually, a given beam profile may correspond to multiple combinations of mode proportion and phase\cite{MD_intensity_distribution_corresponds_multiple_mode_proportions}. However, through repeated GA executions , a unique result of mode proportion and phase is consistently obtained. In other words, the control of the electric field (or phase distribution) of the output beam can be achieved by leveraging its intensity information as feedback. 

\begin{figure*}[th!]
\includegraphics[width=0.95\linewidth]{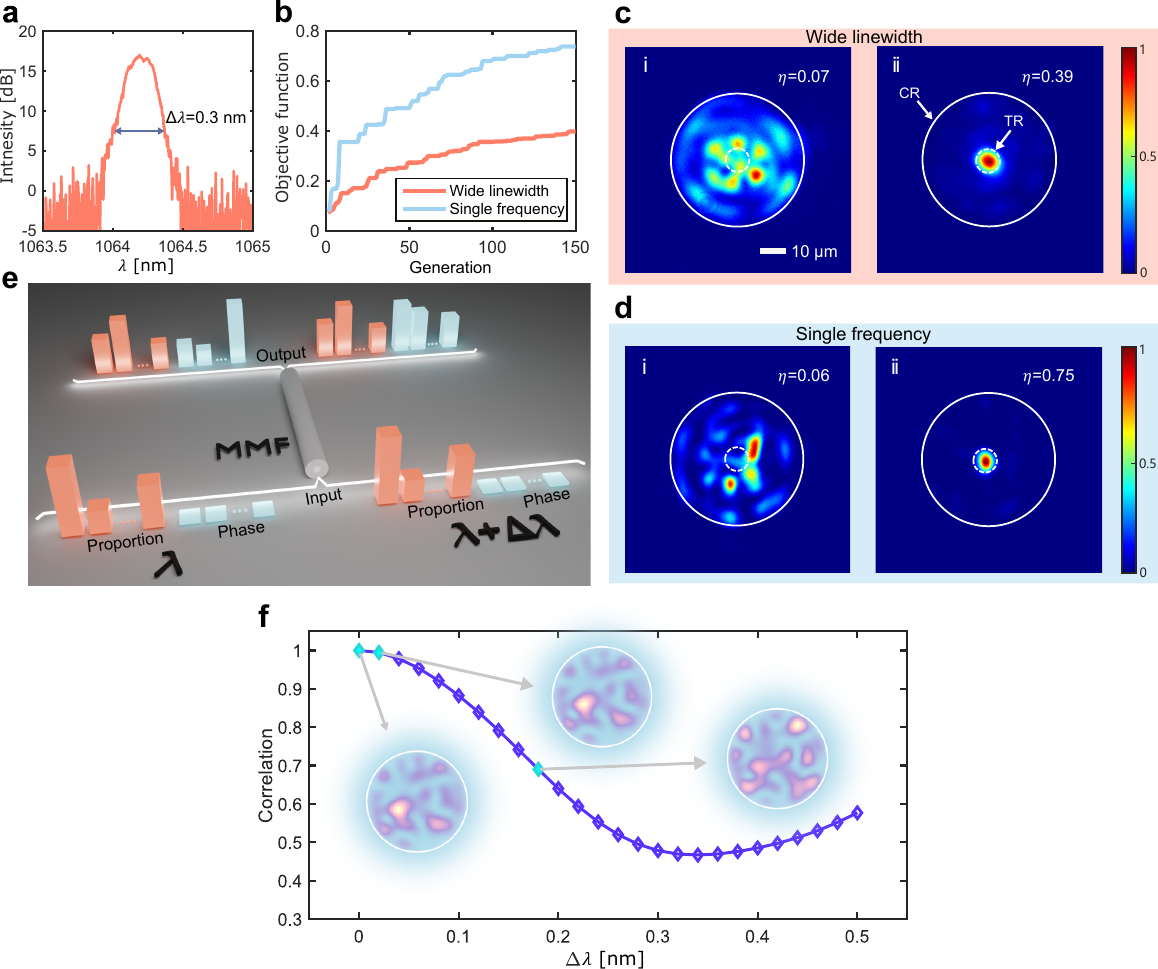}
\centering
\captionsetup{singlelinecheck=off, justification =centerlast}
\caption{{\bf Effect of the linewidth of seed laser on WFS.} {\bf a} The optical spectrum of the broad-linewidth seed. The 10 dB linewidth is measured to be 0.3 nm in the spectrum. {\bf b} The evolution of the objective function during the optimization process. {\bf c} Output beam profiles before(i) and after(ii) optimization with the broad-linewidth seed injected. {\bf d} Output beam profiles before(i) and after(ii) optimization with the single-frequency seed injected. {\bf e} Schematic diagram of the mode proportion and phase for different wavelengths in MMF before and after amplification. Within the laser linewidth range, the output mode proportion primarily depends on mode order, while the output phase depends on both mode order and wavelength. {\bf f} The simulated correlation of the output beam profile at different wavelengths under the same excitation conditions. The insets show the beam profile of 3 chosen wavelengths. The core diameter of MMF used in simulation is 50 $\si{\micro\meter}$ and the NA is 0.12. The propagation distance is set to 2 m. The wavelength sampling interval in the simulation is 0.02 nm. To illustrate the differences in speckle patterns, only the first 50\% of the modes supported by the fiber are excited.
}
\label{fig4}
\end{figure*}

It is worth noting that the entire mode space is explored in the simulation, which constitutes a (2N-1)-dimensional space defined by N dimensions of mode proportion and N-1 dimensions of intermodal phase. This assumes complete control over the electric field at the input facet of the MMF, which requires manipulating both phase and amplitude of the electric field on SLM. In practice, the phase-only SLMs are typically used for their superior power handling capability, which inherently limits its control over the electric field on the input facet of the MMF. Nevertheless, in scenarios where a nearly 100\% coupling efficiency of seed is not required, the optimization criteria can be relaxed such that the optimal proportion and intermodal phase is achieved through the projection of the electric field into the mode space at the fiber input facet.

\vspace{3 mm}
\noindent{\bf 3.Effect of the laser linewidth on the optimization performance}

Most previous WFS studies on MMF were based on monochromatic lasers, whereas in many high-power applications such a narrow linewidth is not necessary. Therefore, we investigated the linewidth of the seed laser required for effective focusing of the output beam in MMF. WFS experiments were conducted utilizing both single-frequency and broad-linewidth seed lasers. The spectrum of the broad-linewidth seed is presented in Fig.~\ref{fig4}a. The single frequency seed operates at 1070 nm with a linewidth of 20 kHz. The MMF employed has a core diameter of 50 $\si{\micro\meter}$ and a NA of 0.12. The length of the MMF is approximately 2 m. Experimental results are presented in Fig.~\ref{fig4}b-d. Notably, the final objective function value for the broad-linewidth seed is approximately half of that with the single-frequency seed, which indicates that for a 2-meter MMF with NA=0.12, the linewidth of the broadband seed laser already exceeds the acceptable tolerance range for WFS.

To elucidate the impact of linewidth on the control of the output beam profile, we first theoretically analyze the determinants of the output beam profile for different wavelengths. It has been illustrated in Fig.~\ref{fig3}c that the global optimum of $\eta$ corresponds to unique mode proportion and phase for each  frequency component of the laser. After being modulated by the SLM,the excited mode proportion and intermodal phase on the input facet of the MMF are identical across different frequency components. However, discrepancies arise in the output states among different frequency components, where the mode proportion depends on gain and the phase depends on inter-modal dispersion. The laser gain is primarily governed by the emission cross-sections of $\text{Yb}^{3+}$ and the overlap integral with the doped region of different modes field. For the wavelength span of most laser linewidth, the discrepancy of gain among different wavelength components is negligible. Therefore, in multimode fiber laser amplifiers, we primarily consider the intermodal phase difference among different frequency components in the output beam caused by inter-modal dispersion (Fig.~\ref{fig4}e). 

The deviation in the output beam profile between different wavelengths is studied by simulation and characterized by the correlation (defined in Method Sec.~4). When the correlation for wavelengths within the linewidth remains sufficiently high, the resulting variations can be deemed negligible. As shown in Fig.~\ref{fig4}f, achieving effective control of the output laser beam profile WFS necessitates a linewidth narrower than approximately 0.1 nm for the 2 m MMF with a NA of 0.12 employed in the experiment, which is consistent with the experimental observations. For any type of step-index MMF, the frequency difference corresponding to the half-maximum of the correlation is inversely proportional to the fiber length and the square of \(\text{NA}\) \cite{goodman2008speckle}.

\vspace{3 mm}
\noindent{\bf Discussions}  

To enhance the thresholds of mode instability and nonlinear effects, MMFs support a greater number of modes can be adopted\cite{Further-Theoretical-Study-on-High-TMI-Threshold-in-Multimode-Systems_APL_2024,High-TMI-Threshold-in-Multimode-Systems_PNAS_2023,High-SBS-Threshold-in-Multimode-Systems_NC_2023,Optimal_Excitation_for_Improving_TMI_and_SBS_Thresholds_in_MMF_2024}. However, fibers with a larger NA require narrower linewidth of the seed laser due to the spectral de-correlation of the output beam profile. Additionally, a larger NA corresponds to a smaller diffraction-limited spot size, which increases the power density at the fiber facet. Fibers with larger core diameters reduce the power density of the signal laser within the core, which demands a seed laser with higher power to suppress ASE.

It should be noted that when moving the CCD out of focus, distortions of the focused spot can be observed. These distortions can be attributed to the weak background that persists within the CR, caused by the limitations of focusing ability in WFS. This residual background interferes with the focused spot in the Fresnel diffraction zone, leading to a high $\mathrm{M}^2$  factor when directly measured utilizing a $\mathrm{M}^2$ analyzer. To address this issue, spatial filtering of the focused laser spot at the output facet of the MMF may be employed, for instance, by employing a pinhole or splicing a short segment of single-mode fiber with CLS(see the Supplementary Sec.~3 for more details). 

Since the output state of a multimode fiber laser is highly dependent on the pump power, the WFS must be conducted at a predetermined high-power level with sufficient pump power. However, the inherent randomness in GA during the WFS process introduces randomness in the coupling efficiency, which should be carefully monitored to avoid ASE. The output beam profile of the system is sensitive to temperature fluctuations. The intense heat dissipation of MMF in high-power amplification regimes and the requirement for temperature stability impose stricter requirements on WCGF, which may be resolved by using faster wavefront modulators (such as deformable mirrors) and high-speed CCDs, since most of the time in the optimization process is consumed waiting for the SLM to respond. Furthermore, a recent simulation study proposed a approach based on spacetime symmetry, which may solve the instability of the system in high-power regime\cite{Theoretical_Study_of_Multimode_Amplifiers_Based_on_WFS_2024}.

Coupling the seed laser at the MMF facet demands high precision in the alignment of the optical path, which requires frequent maintenance and is vulnerable to external vibrations. We propose that an all-fiber version may be achieved by controlling the intermodal phase with proportion uncontrolled, where higher-order modes are excited though heterogeneous splicing (see the Supplementary Sec.~2 for more details). 

In summary, the WFS technique is employed to focus the output beam profile of a multimode fiber laser amplifier while maintaining high laser amplification efficiency, resulting in a 168 W high-brightness multimode fiber laser output. The output beam profile is tailored by carefully selecting the size of the TR. Compared to the multimode beam profile without control, the brightness is improved by approximately eightfold with the same power. These results demonstrate the effectiveness of WFS in controlling the beam profile of multimode fiber lasers in the high-power amplification regime, enriches the architectures of fiber laser and show an innovative way for achieving higher brightness in fiber laser systems.

\vspace{3 mm}
\noindent{\bf Materials and methods}

\vspace{1 mm}
\noindent \textbf{1. Optical setup}

The experimental configuration of the multimode fiber laser amplifier comprises three parts: WFS part, fiber part, and output detection part, as illustrated in Fig.~S1. 
In the WFS part, the polarization of collimated seed laser is rotated to horizontal direction. After being reflected by the reflective phase-only SLM, it is coupled into the MMF through a lens with a focal length of 5~cm. A 9:1 BS (beam splitter) diverts 10\% of the laser to a power meter for monitoring the power of seed laser. 
The fiber part includes a cladding light stripper (CLS), a gain fiber, a 6+1 pump-signal combiner and quartz block head (QBH). The gain fiber is a Yb-doped fiber (YDF 48/400, NA=0.06) with a length of 3~m. The amplifier is backward-pumped by 976 nm laser diodes via a fiber pump-signal combiner. The total length of passive fiber in the multimode fiber laser amplifier is approximately 5~m (GDF 48/400 NA=0.06 ). 
In the output detection part, the amplified output beam is collimated by a lens with a focal length of 3.5~cm. The collimated beam is reflected by a mirror with a reflectivity of 99.9\% into a power meter, while 0.1\% of the transmitted laser is utilized for beam diagnostics. CCD2 captures the intensity distribution at the output facet of the MMF, which is mounted on a single-axis motorized translation stage.
A detailed description of the optical setup is presented in Sec.~1 of Supplementary Information.

\vspace{1 mm}
\noindent \textbf{2. Implementation of GA in WFS experiments}

The basic form of GA is similar to that in Ref. \cite{Proposal_of_GA_in_WFS_2012}. The \(1024 \times 1024\) central pixel of the SLM (\(1272 \times 1024\) pixel with size of \(12.5 \, \mu\text{m}\)) is grouped into \(16 \times 16\) macropixels, striking a balance between optimization performance and the coupling efficiency of the seed laser. The gene encoding of each individual (phase map) in GA is the integer values of 256 macropixels on the SLM. Each macropixel value is an integer ranging from 0 to \(212\), corresponding to a phase modulation range of \(0\) to \(2\pi\) of the SLM at a wavelength of 1045~nm, resulting in \(212^{256}\) possible phase maps. The number of individuals per generation is set to 30. Offspring are generated by selecting parents from the previous generation using a roulette wheel selection method based on their objective function values. A random mutation is applied to certain macropixels according to a predefined mutation probability. The mutation probability is initially set to 0.3 and decreases exponentially with successive iterations (see the inset of Fig.~S1). The offspring, constituting 50\% of the population, replace the least fit individuals from the previous generation. The number of generations is set to 100–150. The objective function of GA is defined as power ratio with the TR. When the TR is positioned at the center of the fiber core, the majority of the output power is distributed within the first 15 modes (as shown in Fig.~2c), therefore, the divergence angle of the output beam should be smaller than the NA of the fiber, which depends on the normalized propagation constant of the highest-order mode. Consequently, the actual size of diffraction-limited spot after optimization can be calculated by $1.22\frac{\lambda f}{d}$ though the measuring the diameter $d$ of collimated output beam. Before optimization, the final mode proportion is unknown exactly, therefore the size of TR is simply calculated by \( \frac{\lambda}{\pi \times \text{NA}} \).

Each new phase map is loaded on the SLM, and after a 150~ms response time for the liquid crystal molecules, the objective function is calculated according to the beam profile captured by CCD2. The complete optimization process for the GA takes approximately 5–8 minutes. The individual with the highest objective function in each generation is used to plot the evolution of the objective function and the corresponding output beam profile. It was observed that permitting change the position of the TR during the later stage of optimization will reduce the convergence time. Consequently, the position of the TR was allowed to track the already-formed focal spot after the 70th generation, which resulted in the final spot slightly deviating from the central position.

\vspace{1 mm}
\noindent \textbf{3. Implementation of GA in WFS simulations}

The GA encodes individual genes as \((2N-1)\)-dimensional vectors in the mode space, consisting of \(N\) mode proportion and \(N-1\) intermodal phase. The objective function is defined as the power ratio of the output beam profile within the target region (TR). The mode fields $\psi_m \left( x,y \right)$ and the electric field of beam profile ${E}(x, y, z)$ obtained through modes superposition are power-normalized:
\begin{equation}  
\begin{gathered}  
{E}(x, y, z) = \sum_{m} A_m {\psi}_m(x, y) e^{i\beta_m z+\phi_m} + \text{c.c.}\\\
\sum_{m} |A_m|^2=1  \\
\iint{\frac{1}{2}c}\varepsilon_0 n_{core}\psi_m \left( x,y \right) \psi_m ^*\left( x,y \right) \mathrm{d}x\mathrm{d}y=1
\end{gathered}  
\end{equation}
Where $\beta_m$, $|A_m|^2$ and $\phi_m$ are the propagation constant, proportion and phase of mode $m$, respectively. $c$ is the speed of light, $\varepsilon_0$ is the vacuum permittivity and $n_{core}$ is the refractive index of fiber core. The mutation probability parameter defines the number of elements mutating in the offspring, which is set to 0.5 in the simulation. The mutation magnitude of the proportion and phase are defined respectively. When the mutation amplitude is set to 1, the variation range for mode proportion is \(0\) to \(1\), while the phase variation range is \(0\) to \(2\pi\). In the simulation, the mutation amplitude for mode proportion is adjusted according to the number of fiber modes, while the mutation amplitude for phase is uniformly set to \(0.2\), and both mutation amplitudes decrease with the increase of generations.

\vspace{1 mm}
\noindent \textbf{4. Definition of correlation}

The correlation between two beam profiles is quantified by pearson correlation coefficient (4), where $I_0(x,y)$ represents the reference beam profile captured at the initial moment, averaged over 30 frames, and $I(x,y,t)$ denotes the real-time output beam profile. The correlation $C(t)$ quantifies the stability of the beam profile over a period of time.
In the simulated beam profile of different wavelengths, the $t$ is replaced by the wavelength discrepancy $\Delta \lambda$. And the correlation $C(\Delta \lambda)$ quantifies the spectral decorrelation in the beam profile.
\begin{equation}
\begin{gathered}  
C(t) = \frac{\iint \Delta I_0(x,y,0)\, \Delta I(x,y,t)\,\mathrm{d}x\,\mathrm{d}y}  
{\sqrt{\iint \left[\Delta I_0(x,y,0)\right]^2 \,\mathrm{d}x\,\mathrm{d}y   
       \iint \left[\Delta I(x,y,t)\right]^2 \,\mathrm{d}x\,\mathrm{d}y}},\\
\Delta I_0(x,y,0) = I_0(x,y,0) - \bar{I}_0(0),\\
\Delta I(x,y,t) = I(x,y,t) - \bar{I}(t)  
\end{gathered}  
\end{equation}

\vspace{1 mm}
\noindent \textbf{Acknowledgments}

We gratefully acknowledge supports from the Science and Technology Innovation Program of Hunan Province (2021RC4027) and the National Natural Science Foundation of China(62405373).

\vspace{1 mm}
\noindent\textbf{Contributions} 

Zefeng Wang conceived the idea. Zhen Huang, Binyu Rao conducted the experiments. Zhen Huang and Chenxin Gao conducted the simulation. All authors discussed the results and contributed to writing the manuscript.

\vspace{1 mm}
\noindent \textbf{Data availability}

The data underlying the results presented in this paper and the code of the optimization algorithm are not publicly available at this time but may be obtained from the authors upon reasonable request.

\vspace{1 mm}
\noindent \textbf{Conflict of interest} 

The authors declare no conflicts of interest. 


\vspace{1 mm}
\bibliography{ref}

\end{document}